\documentclass[twocolumn,aps,prb,showpacs,floatfix]{revtex4}
\usepackage{graphicx, graphics, color}

\input amssym.tex
\input amssym.def

\newcommand{\be}{\begin{equation}}
\newcommand{\ee}{\end{equation}}
\newcommand{\ben}{\begin{eqnarray}}
\newcommand{\een}{\end{eqnarray}}

\begin{document}

\title{Holographic superconductivity in the presence of dark matter: basic issues}

\author{Marek Rogatko} 
\email{marek.rogatko@poczta.umcs.lublin.pl }
\author{Karol I. Wysoki\'nski}
\email{karol@tytan.umcs.lublin.pl}
\affiliation{Institute of Physics \protect \\
Maria Curie-Sklodowska University \protect \\
20-031 Lublin, pl.~Marii Curie-Sklodowskiej 1, Poland}

\date{\today}

\begin{abstract}
The holographic approach to study strongly coupled superconductors in the presence
of dark matter is reviewed. We discuss the influence of dark matter on the
superconducting transition temperature of both s-wave and p-wave holographic 
superconductors.  The upper critical field, coherence length, penetration 
depth of holographic superconductors as well as  the
metal-insulator transitions have also been analysed.
Issues related to the validity of AdS/CFT 
correspondence for the description of superconductors studied in the laboratory and   
possible experiments directed towards the  detection of dark matter
 are discussed. In doing so we shall compare our assumptions 
and assertions with those generally accepted in the elementary particle experiments 
aimed at the detection of dark matter particles.
\end{abstract}

\maketitle
\section{Introduction}
The superconductivity is a macroscopic quantum phenomenon~\cite{tinkham2004} appearing in many condensed matter
systems and characterized by the expulsion of the magnetic field and zero resistance state at sufficiently
low temperature. 
The  superconducting rings show quantization of magnetic flux. This is also true
for vortices in the type two superconductors.  A number of other  
quantum  effects appear if the superconductor is in contact with normal metal (Andreev reflection) 
or other superconductors (Josephson effect). 
All these properties and phenomena are at the heart of numerous applications of superconductors.

Zero resistivity state allows the construction and operation of powerful superconducting electromagnets
for important scientific and medical applications. The superconducting quantum interference devices (SQUIDs)
are the most sensitive magnetometers. SQUIDs allow for the precise measurements of the flux quantum 
$\phi=h/2e$, where $e$ is the electron charge and $h$ the Planck constant. These devices  have 
been proposed and used in the search for the magnetic monopole; the elementary particle 
~postulated\cite{dirac1931} by Dirac 
in 1931. The search~\cite{cabrera1982} for this elusive particle continues~\cite{burdin2015},
with the hope that it might address an important question  of composition of dark matter~\cite{courteau2014}.

According to standard theory of gravity based on Newtonian dynamics the explanation of many astrophysical 
observations require the existence of large amount of matter that cannot be seen with telescopes, and is 
thus termed {\it dark matter}. 
The Planck satellite mission reveals that {\it dark matter} constitutes 26.8\% of the total mass 
of the Universe, while ordinary matter makes only 4.9\% of it.
The rest~\cite{ade15} is the mysterious dark energy, 68.3\%.

The nature of the dark matter is still unknown. Besides magnetic monopoles
other particles have been proposed as possible candidates. 
Many experiments~\cite{hooper2007} aimed at the direct detection of the postulated dark matter 
particles have been proposed.  Some of them~\cite{beck2013,wilczek2014} rely on 
the use of superconductors and study of subtle quantum (interference) effects.   
We shall not discuss various proposals for the dark matter particles~\cite{bertone2005} here. 
Instead we model the dark matter classically by the additional field with 
properties similar to the Maxwell field and study  the  influence of dark matter on
 the properties of holographic  superconductors. The main goal of the present work is 
to discuss how the presence of  dark matter may change the properties of holographic 
superconductors. Assuming that the holographic analogy teaches us about the strong 
coupling aspects of superconductors studied in the laboratory we discuss possible   
experiments directed towards the detection of this elusive but dominant part of matter 
in the Universe. 

The rest of the paper is organized as follows. We discuss the main ideas of the general holographic approach
to study strongly coupled models in Section II, where we also briefly state the main effects of
the dark sector on the properties of superconductors. In Section III we shall discuss general questions 
of matter composition and the geometry of the Universe, chances of  direct detection of dark matter field in the
laboratory by long time observations of superconductors. We also offer some remarks on the 
possibility of superconducting condensation of dark matter. We end up with conclusions.

\section{The effect of dark matter on superconductors within AdS/CFT correspondence}
\begin{figure}
\includegraphics[width=0.95\linewidth]{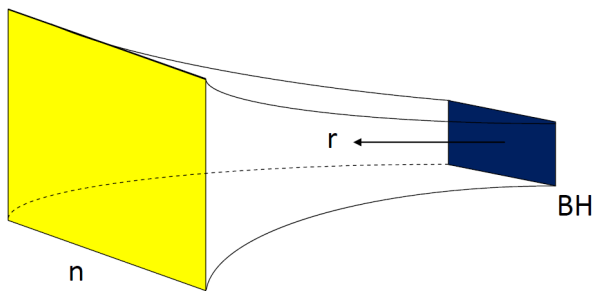}
\caption{In the holographic approach the shadowed plane on the right represents the
black hole (BH). The extra 'radial' dimension is denoted by \textbf{r}. The light 
shadowed plane on the left represents the boundary $\textbf{r}\rightarrow \infty$
of the AdS $n$-dimensional space-time, $i.e$ for n=5 the Minkowski space ($t,x,y,z$)
of the field theory. The region of finite \textbf{r} ($0<\textbf{r}<\infty$) is denoted
as gravitational bulk. All processes in the bulk represent long distance or low energy processes (IR limit),
while those close to boundary correspond to short distance or high energies (UV limit). Thus
distance \textbf{r} correspond to the renormalisation group parameter deciding
about the coupling strength.}
\label{fig1}
\end{figure}

The gauge/gravity duality is also known as holographic duality
or anti-de Sitter/conformal field theory  (AdS/CFT) correspondence~\cite{mal99}.  It is the conjectured
equivalence between the field theory in the Minkowski space-time forming the
boundary of the Einstein -- anti-de Sitter (AdS) and the gravity theory in the bulk of  
the AdS space-time, see Fig. \ref{fig1}. There exist a kind of a 'dictionary' which relates the
results of calculations in one theory to those obtained or expected in the other.
In other words, every property in one theory has a counterpart in the other theory. 
In order to study temperature dependent phenomena one assumes the existence 
of black hole in the gravity theory. The AdS/CFT correspondence is typically viewed as
a prescription how to calculate the expectation values of the operators 
in the strongly coupled field theory by means of the gravity theory. The latter
is typically easy endeavor as the gravity action is dominated by classic
field configurations. It is believed that the correspondence provides the
strong coupling - weak coupling duality~\cite{sachdev2012}. In technical terms~\cite{green2013} 
the correspondence is related to the fact, that the partition function of the field 
theory $Z[A({\bf x},t)]$, where  $A({\bf x},t)$ is the source field that couples
to the currents ${\bf j}({\bf x},t)$, equals to that of the gravity dual 
$Z[A({\bf x},r,t)]$ with the boundary condition $\lim_{r\rightarrow \infty}A({\bf x},r,t) =A({\bf x},t)$. 
As mentioned the latter partition function  is prevailed by the classical field configurations of gravitational field
and thus given by the exponential of the classical action of the aforementioned configurations.

The first successful description of  strongly coupled superconductors
by gauge/gravity duality~\cite{hartnoll2008} relied on the planar Schwarzschild –- anti - de Sitter black hole
and Maxwell electrodynamics. The scalar complex field $\Psi(r)$ is 
coupled to the Maxwell field represented by the time component of the four-vector
$A_t=\Phi(r)$. It has to be stressed that properties of the dual field theory can be read off from
the asymptotic behavior $r\rightarrow \infty$  of the field equations. 
In particular the scalar field behaves close to the boundary $r\rightarrow\infty$ as
$\Psi(r)=<O_1>/r+<O_2>/r^2+...$, where according to the general prescription
$<O_1>$ and $<O_2>$ play a role of the condensate density in the dual theory and can
be compared with superconducting order parameter. Both $<O_1>$ and $<O_2>$ take on the non-zero value 
at temperatures (provided by the black hole) lower than the characteristic critical 
temperature $T_c$.  The temperature dependence of $<O_2>$ close to $T_c$
shows the behavior characteristic for the superconducting order parameter $\Delta$ with
the mean-field type dependence $\propto(1-T/T_c)^{1/2}$ close to $T_c$. 

{\it Strong coupling effects in holographic superconductors.} The calculated~\cite{hartnoll2008} ratio
$2\sqrt{<O_2(T=0)>}/T_c\approx 8$ is characteristic for strongly coupled superconductors.
The other strong coupling effects include the upward curvature seen in the temperature dependence 
of the upper critical field calculated {\it via} AdS/CFT correspondence~\cite{nak14,nak15,nak15a,rog2015,rog2015a}.
It leads to the value of zero temperature magnetic 
field higher than the estimate  based on the equation~\cite{klemm1975} 
${B_c}(0)\approx 0.69~ T_c\left(\frac{d{B_c}(T)}{dT}\right)_{T_c}$ relating upper critical field
at $T=0$ to its derivative close to $T_c$. 
The upward curvature, which is observed in many high temperature superconductors in the standard
approach~\cite{klemm1975} requires {\it inter alia} strong potential or spin-orbit scattering, while
it seems to appear quite naturally in the strong coupling.

{\it Generalizations.} Since the first  attempt to apply 
the gravity to construct holographic superconductors~\cite{hartnoll2008},  
there appeared a number of papers generalizing 
the approach in various directions. Among them one has to mention 
the complex gravity backgrounds including higher order curvatures~\cite{gb}, non-linear
Born-Infeld electrodynamics~\cite{binf}, etc. Important results have been obtained by
changing the chemical potential $\mu$, which in the gauge/gravity theory is given by   
the asymptotic behavior of the Maxwell field $\Phi=\mu-\rho/r+...$, where $\rho$ is charge
density. It has been found that increase of  $\mu$ at $T=0$ results in the insulator to metal
transition, which in the gravity theory is related to Hawking-Page transition~\cite{hp} between black soliton
and black hole. Also superconductors with symmetries other than s-wave have been 
studied~\cite{cai2015}.  

{\it Dark matter and AdS/CFT.} Of special interest are the papers studying 
the effect of {\it dark matter} field on such 
properties of superconductors as transition  temperature, the condensate amplitude, upper critical field, 
coherence length, penetration depth, etc.
One assumes that besides the Maxwell field there exist another U(1)-gauge field, which represents that
part of the matter in the Universe, which is not visible with telescopes.
  
The minimal gravitational action  in $n$-dimensional space-time  reads
\be
S_{g} = \int \sqrt{-g}~ d^n x~  \frac{1}{2 \kappa^2}\bigg( R - 2\Lambda\bigg), 
\ee
where $\kappa^2 = 8 \pi G_{n}$ is an n-dimensional gravitational constant.
The cosmological constant is given by $\Lambda = - \frac{(n-1)(n-2)}{2L^2}$,  
where $L$ is the radius of the AdS space-time.
The matter action contains the Abelian-Higgs sector coupled to the second $U(1)$-gauge field.
It is provided by the following action:
\ben
\label{s_matter}
S_{m} = \int \sqrt{-g}~ d^nx  \bigg( 
- \frac{1}{4}F_{\mu \nu} F^{\mu \nu} - \left [ \nabla_{\mu} \psi - 
i q A_{\mu} \psi \right ]^{\dagger} \nonumber \\
\left [ \nabla^{\mu} \psi - i q A^{\mu} \psi  \right ]
- V(\psi) - \frac{1}{4} B_{\mu \nu} B^{\mu \nu} - \frac{\alpha}{4} F_{\mu \nu} B^{\mu \nu}
\bigg). 
\een  
The scalar field potential representing the superconductor satisfies
$V(\psi) = m^2 |\psi|^2 + \frac{\lambda_{\psi}}{4} |\psi|^4$.
$F_{\mu \nu} = 2 \nabla_{[ \mu} A_{\nu ]}$
stands for the ordinary Maxwell field strength tensor, while
the second $U(1)$-gauge field $B_{\mu \nu}$ is given by  
$B_{\mu \nu} = 2 \nabla_{[ \mu} B_{\nu ]}$. Moreover, $m,~ \lambda_{\psi},~ q$ represent
mass,  coupling constant and charge related to the scalar field $\psi$, respectively.
On the other hand, $\alpha$ is a coupling constant between $U(1)$ fields. Analysis
of the three-dimensional superconducting systems living in a 3+1 dimensional space-time, 
requires $n=5$ dimensional gravity theory.

{\it Analogies and differences.}  For condensed matter specialists it is important to note that the field $\Psi$
plays here a very similar role as (usually denoted by the same symbol) field
in the Ginzburg-Landau approach. There is, however, a very important difference.
While the condensation in Ginzburg-Landau approach requires non-zero value of  
the quartic parameter $\lambda_{\psi}$, in the gravity theory the field $\Psi$
undergoes condensation for $\lambda_{\psi}\equiv 0$. The other crucial difference 
has to do with the mass parameter $m^2$. Its counterpart in Ginzburg - Landau approach is usually
denoted by $a$ and is assumed to change sign at the superconducting transition temperature:
$a=a'(T-T_c)$ taking on negative value in the superconducting state.

In a formal analogy, in the gravity approach  $m^2<0$ in the superconducting state.
However, the stability of the AdS $n+1$ dimensional space-time requires fulfilling 
the Breitenlohner-Freedman bound  $m^2 > - n^2/4$. It means the lower limit on the 
the 'mass' term in holographic superconductor. 
While it is generally accepted that the scalar field condenses~\cite{bri11} for $m^2<0$, there are papers
proving the existence of the solution for $m^2>0$ in specific conditions.
Kim {\it et al.} have shown~\cite{kim09} that the holographic superconductor may still exist for the case $m^2 > 0$.
They found that if the mass of the scalar field increases,  the system undergoes 'the phase space 
or configuration space folding', i.e., the two very close boundary conditions 
can lead to the different states. For instance, one with the broken symmetry while the other is a symmetric one.

The other tantalizing question is the inquire about the meaning of the limit for 
superconducting material, if any? 
Would its existence translate to the 
the  limit of the superconducting gap function? In the  BCS approach 
the magnitude  of the gap function is directly proportional to the electron-phonon
coupling constant. However, the very large values of the coupling constant make
the BCS weak coupling approach not valid. The Eliashberg method usually considered as
a strong coupling theory, in fact takes retardation effects into account and is
valid beyond BCS limit. However, based on the expansion with respect to electron-phonon
coupling approach the theory is not a strong coupling one and possible instabilities at larger
couplings have been analyzed and predicted. The detailed analysis of the relation between
Breitenlohner-Freedman bound and the gaps in the field theory would shed
additional light on the superconducting instability of materials, but this is outside 
the scope of the present paper.

{\it Some results.} Studying the influence of dark matter on the properties of superconductors
we are mainly interested in the changes of their properties as characterized
by the coupling constant $\alpha$ (see Eq. \ref{s_matter}). 
Working  in the probe limit we have found that dark matter sector affects various properties 
of the holographic  superconductors depending on their symmetry and other details.
In particular the superconducting  transition temperature $T_c$ of the s-wave
superconductor with dark matter (DM) is modified as $T_c^{DM}=T_c+\tilde{\alpha}T_0$, where
$T_c$ is the superconducting transition temperature without dark sector and $T_0$ is constant depending on the
dimensionality of space-time and other model parameters, $\tilde{\alpha}=1-\alpha^2/4$. 
Similar dependence for the model of SU(2) 
Yang-Mills p-wave holographic superconductor reads  $T_c^{DM}=T_c/\tilde{\alpha}^{1/6}$, while
in the the Maxwell vector p-wave holographic superconductor
with dark matter sector and for real  components of the vector field give us the same description
of the phase transitions as the s-wave model~\cite{nak14,nak15,nak15a,rog2015,rog2015a}. It is worth noting the  
dependence of the superconducting  transition  temperature $T_c$ 
 on the  charge density $\rho$, which  is $T_c\propto \rho^{1/3}$. The value $1/3$ of the power
factor in a three dimensional superconductor seem to be strong coupling modification of the
exponent 2/3 known from the Bose - Einstein condensation of charged local pair bosons 
in narrow band superconductors~\cite{micnas1990}.
The coherence length $\xi$ and the penetration depth $\lambda$ both feature the 
dependence on the coupling $\alpha$ with $\lambda^{DM}=\lambda \sqrt{\tilde{\alpha}}$ and
$\xi^{DM}=\xi/\sqrt{\tilde{\alpha}}$. As a result the presence of dark matter influences
both length scales and the parameter $\kappa^{DM}=\tilde{\alpha}\kappa$. As $\kappa^{DM}$ may decrease
with increasing $\alpha$ the superconductor may  change~\cite{rog2015} from the weakly 
second type (with $\kappa^{DM}>1/\sqrt{2}$) to first kind ($\kappa^{DM}<1/\sqrt{2}$). If observed, this
change~\cite{misprint} could be an important fingerprint of the effect of dark matter on real superconductors
and not only on the holographic ones. 
Most of the above results have been obtained in the probe limit, see however~\cite{peng2015} 
for recent analysis of the effect of dark sector with back-reaction taken into account.

\section{Miscellaneous issues}
Very often we are asked about the motivation
why to study the influence of a dark matter sector in holographic superconductors. 
Typically interlocutors admit that dark matter may play an important role in gravitational
theories of the Universe and inquire: if it is important for superconductors;  if and how the dark matter might
be detected in the laboratory by studying real superconductors.
 Another important aspect of discussion is related to the fact that our Universe
has Einstein - de Sitter geometry, but hypothetical  anti- de Sitter space-time is assumed
in the gauge/gravity correspondence. 
The above problems  are at the heart of the gauge/gravity duality.  

{\it Relation to real life systems.} The first question is whether the superconductors studied 
in the holographic theory have something to do with real life systems studied in
the lab. We tend to believe that the AdS/CFT analogy describes or at least gives hints
 about the behavior of  real systems.
We have the following arguments in favor of this conviction:

(i) The duality has helped to understand  
the extremely low viscosity of the quark-gluon plasma. While perturbative approaches~\cite{shuryak2004}
fail to provide the understanding of the puzzle, the gauge-gravity duality
finds the universal scale for the viscosity~\cite{kovtun2005}.

Applied to study superconductors 
the  gauge-gravity duality describes the behavior expected to be observed in real life 
superconductors (phase transition, critical exponents, etc.). That it provides  sensible and  not 
random answers for various questions is an important argument in  favor of its relevance to real world.
We mention here but a few examples.

(ii) It shows the appearance of condensation at a given temperature with correct or close to correct
temperature dependence
of the parameter expected to describe the condensate~\cite{hartnoll2008} and the expected large
value of the ratio between the gap and the superconducting transition temperature.

(iii) As a further example we quote a  very nontrivial predictions of the 
AdS/CFT theory which {\bf do agree} with other (quantum critical) models~\cite{she2011}.  In that work  
the calculations of the dynamical pair susceptibility by Zaanen and collaborators 
 are presented. Moreover the predictions are expected to be measured in the near future. 

 (iv) The analogy seem to provide deep understanding of the relations (scaling properties) observed near
quantum critical points as observed in real materials.
This is demonstrated e.g. in recent studies of incoherent transport in clean metals~\cite{davison2015}.

Summarizing this aspect,  it seems   
 that the AdS/CFT is saying something new about condensed matter problems in general and  
 holographic superconductors in particular and this is encouraging. We accept the arguments 
of P. W. Anderson  that many properties of superconductors~\cite{anderson2013} are "overdetermined
by experimental facts" and that not every property has been adequately answered by the AdS/CFT duality. 
However, we think that the discussed applications provide an opportunity for field theories by finding 
solutions to strong coupling problems. At the same time the calculations provide real life applications 
of string theory ideas.

{\it AdS/CFT vs. dS Universe.} The relation between the description of the system 
possessing conformal property by the field theory 
and  by the gravity dual is really a tricky one and all the objections may eventually turn out well founded. 
Our understanding of the gauge-gravity duality is the following. Embedding the field theory 
into the AdS space time with the ordinary matter enables us to calculate the properties of the
system (which the field theory is supposed to describe) in the strong coupling~\cite{sachdev2012}
by perturbative methods of the gravity theory evaluated at the boundary. 
Thus we treat/consider the AdS/CFT correspondence as a kind of  
calculus, which enables relatively easy access to strong coupling limit of the CFT.

Using the AdS/CFT approach and studying the properties of holographic superconductors
in the background with dark matter sector we rely on the following.
First we accept that the gauge-gravity duality teaches us about strongly coupled 
superconductors which are produced and studied in the laboratories. Some examples
of this can be found in~\cite{green2013}. The AdS/CFT correspondence relies on the AdS space time. 
Its validity to perform the calculations is not related 
whether the Universe in which the superconductors exist itself forms the  AdS or de Sitter 
space-time. The dark matter existence seem to be obvious from astrophysical data as discussed
in the Introduction. If it interacts with ordinary matter as 
proposed  in  Eq. (\ref{s_matter}) and quantified by the coupling constant $\alpha$, 
so it modifies the behavior of the ordinary matter. These
modifications are found to influence the superconducting transition temperature
and could, in principle, be observed as the annual changes of the properties
of superconductors  following the expected annual changes in the distribution of the
dark matter~\cite{freese2013}.

{\it Dark matter.} Finally we discuss  the relation between
the dark matter as required in astrophysics and its effect on the behavior of 
superconductors. The very notion of dark matter is an old one. The term first appeared in the paper
by Koteyev and is used ever since~\cite{frenk2012}.  Our understanding of the problem
 relies on the following arguments, 
supported by the observations and related work aimed at the detection of dark matter. 
There exist numerous astronomical observation which are described in terms of dark matter.
The precise nature of it is not well known. As already mentioned different  particles
have been proposed as the candidates.  
If the interpretation of astronomical observations in terms of dark matter is true, 
so dark matter must exist and, through its coupling - here $\alpha$ - will modify the properties
of the ordinary matter and thus also the properties of superconductors.
We try to pinpoint, how it may change the properties of the superconductors. If the
calculated changes of the holographic superconductors are indeed expected to apply
to real materials than there is a chance to observe these changes and thus    
get an information on the local dark matter distribution.

The most important question of the experimental Colleagues is related
to the feasibility of the  detection of the dark matter 
in the laboratory. We all accept the fact that the dark matter exists over there
in Cosmos. But if the dark matter exists, as required by the astrophysical
observations,  so it is present not only in the macrocosm, but also in
our neighborhood and may be spotted during the annual motion of the Earth~\cite{freese2013}.
The possible influence  of the dark matter on superconductors can 
in principle be detected by precise and cleverly designed experiments looking at the
annual changes of their properties. We rely here on the arguments presented in~\cite{freese2013}, 
where the authors analyze the annual modulations of the dark matter.
Our additional assumption is that dark matter is non-homogeneously distributed in the neighborhood 
of the Sun~\cite{jungman1996} and these inhomogeneities will make a difference.

The gravity is a classical field theory. It means that there are no particles in it. 
The electromagnetism is described by the Maxwell theory. In the same manner we describe
the dark matter as another classic field, analogous to the electromagnetic one. 
The same model we are using here as described in the previous section, has been 
recently studied in the context of particle physics~\cite{davoudiasl2012}.
These authors hope to observe the signatures of dark matter in the 
 ``planned polarized electron scattering experiments''.
The hope is that polarized electrons in those experiments will respond 
differently because of the dark matter. In the same mood,
 we do hope that electrons in superconductors will also 'feel' the dark sector. 
 
Another question is related to the possible condensation of the dark matter itself. Indeed,
the model with both matter and dark matter vortices has been  recently studied~\cite{ariasa2014},
but the discussion of the results  is outside the scope of the present paper.

\section{Conclusions}	
We have discussed the general approach recently developed to study strongly coupled superconductors
by means of the theory of gravity. The main goal of our study is to find the
influence of dark matter on the properties of superconductors. In the approach the signatures
of dark matter are visible as $\alpha$ dependence of various properties 
of superconductors. In the probe limit, when one neglects the back-reaction of the
condensing field on the metric, the coupling strength $\alpha$ enters various parameters in a multiplicative 
way making it difficult to identify the presence of dark matter. 

Combined with the evidence
of dark matter in the inner Milky Way,  its non-homogeneous distribution
around the Sun  provides a hope of  dark matter 
detection in the precise experiments with superconductors. Similarly to the 
particle experiments aimed at the detection of dark matter using the fact that ``The count rate 
should experience an annual modulation due to the relative motion of the Earth 
around the Sun''~\cite{freese2013} we also expect that the properties of 
superconductor will show annual modulation.
  
The application of the string theory to condensed matter problems is relatively
recent endeavor. Despite that, many novel results have been obtained. However, much more
has to be done. 	
Let us finish  with the following cautionary statements.  
 The field of holographic superconductivity is a developing one and not all relevant questions have 
been asked and adequately answered. Similarly, the concept of dark matter requires 
a lot of theoretical studies and observations before it 
will be properly and completely understood. 
Our work~\cite{nak14,nak15,nak15a,rog2015,rog2015a} is directed towards better understanding of both: strongly
coupled superconductors and seemingly omnipresent, but still illusory dark matter.

\acknowledgments
MR was partially supported by the grant of the National Science Center 
DEC-2013/09/B/ST2/03455 
and KIW by the grant DEC-2014/13/B/ST3/04451.




\begin{thebibliography}{99}
\bibitem{misprint} We take the opportunity and correct the missprint in recent
paper~\cite{rog2015} which erroneously concludes that $\kappa$ is unaffected by the
dark matter.
\end{thebibliography}
\end{document}